\documentstyle[aps,twocolumn,epsfig]{revtex}
\begin{document}
\draft
\title{Nonleptonic $\Omega^{-}$ decays and the Skyrme model}
\author{G. Duplan\v{c}i\'{c}$^1$, H. Pa\v{s}agi\'{c}$^2$, M. Prasza{\l}owicz$^3$,
and J.Trampeti\'{c}$^1$}
\address{$^1$Theoretical Physics Division, R. Bo\v{s}kovi\'{c} Institute, \\
P.O.Box 180, 10002 Zagreb, Croatia\\
$^2$Faculty of Transport and Traffic Engineering, University of Zagreb, \\
P.O. Box 195, 10000 Zagreb, Croatia\\
$^3$M. Smoluchowski Institute of Physics, Jagellonian University, \\
Reymonta 4, 30-059 Krak{\'o}w, Poland}
\date{\today}
\maketitle

\begin{abstract}
Nonleptonic $\Omega^{-}$ decay branching ratios are estimated by means
of the QCD enhanced effective weak Hamiltonian supplemented by the SU(3)
Skyrme model used to estimate the nonperturbative matrix elements. The
model has only one free parameter, namely the Skyrme charge $e$, which is
fixed through the experimental values of the octet-decuplet mass splitting $%
\Delta $ and the axial coupling constant $g_{A}$. The whole scheme is
equivalent to that which works well for the nonleptonic hyperon
decays. The ratios of calculated amplitudes are in agreement with experiment.
However, the absolute values are about twice too large if short-distance corrections
and only ground intermediate states are included.
\end{abstract}

\pacs{PACS number(s): 12.39.Dc, 12.39.Fe, 13.30.Eg}

\narrowtext
Recently both s- and p-wave nonleptonic hyperon decay amplitudes were quite
successfully reproduced by the SU(3) extended Skyrme model with the QCD
enhanced effective weak Hamiltonian \cite{du}. The decay amplitudes were
described through the current-algebra commutator, the ground-state baryon
pole terms and factorizable contributions. The nonperturbative quantities,
{\em i.e.} the baryon 4-quark operator matrix elements, were estimated
using the SU(3) Skyrme model. For the s-wave hyperon decay
amplitudes correct relative signs and absolute magnitudes were obtained. For
the p-waves all relative signs were correct, with their relative
magnitudes roughly following the experimental data. As far as the absolute
magnitudes are concerned, the poorest agreement between theory and
experiment was within a factor of two. One is thus faced with an obvious
question: could an analogous approach work equally well for the $\Omega ^{-}$
nonleptonic weak decays?
Such a question should be considered in connection with
the accurate measurements of the $\Omega ^{-}$ decay branching ratios.
The experimental value for the $\Omega ^{-}$ mean life time is \cite{rpp}:
\[
\tau _{\rm experiment}(\Omega ^{-})=(82.1\pm 1.1)\;\; {\rm ps}
\]
from which the values of the p-wave amplitudes given in Table \ref{t:tab1}
have been extracted.

In the Skyrme model, baryons emerge as soliton configurations of the field $%
U $ of pseudoscalar mesons \cite{sky,adk}. Extension of the model to the
strange sector \cite{gua,NMPJW,MPSU3,wei} is done by an isospin embedding of
the static {\em hedgehog} Ansatz into an SU(3) matrix which is a subject of
a time dependent rotation
\begin{equation}
U(\vec{r},t)=A(t)\left(
\begin{array}{cc}
exp(i{\vec{\tau}\cdot }\vec{n}F(r)) & 0 \\
0 & 1
\end{array}
\right) A^{\dagger }(t),  \label{U}
\end{equation}
by a collective coordinate matrix $A(t)\,\in \,SU(3)$ which defines the
generalized velocities $A^{\dagger }(t)\dot{A}(t)={\frac{i}{2}}\sum_{\alpha
=1}^{8}\lambda _{\alpha }\dot{a}^{\alpha }$ and the profile function $F(r)$
is interpreted as a chiral angle that parametrizes the soliton. The
collective coordinates $a^{\alpha }$ are canonically quantized to
generate the states that possess the quantum numbers of the physical strange
baryons. In order to account for a non-zero strange quark mass the
appropriate chiral symmetry breaking (SB) terms should be included. In this
work, however, following \cite{du} we shall neglect the SB effects since
they are much smaller than the uncertainties coming from other sources (like
for example the values of the $c_{i}$ coefficients in the effective
Hamiltonian, or factorization. Moreover the SU(3) symmetry
is used in course of the calculations of the decay amplitudes, so for
consistency the nonperturbative matrix elements have to be evaluated in the
same approximation scheme.

Our goal is to see whether the effective weak Hamiltonian and the SU(3)
Skyrme model are able to predict the nonleptonic $\Omega ^{-}$ decay
amplitudes $(\Omega _{K}):\Omega ^{-}\rightarrow \Lambda K^{-}$, $(\Omega
_{-}^{-}):\Omega ^{-}\rightarrow \Xi ^{0}\pi ^{-}$ and $(\Omega
_{0}^{-}):\Omega ^{-}\rightarrow \Xi ^{-}\pi ^{0}$ following the method of
Refs.\cite{du,ga}. To this end we shall employ the Standard Model effective
Hamiltonian and {\it the minimal number of couplings concept} of the Skyrme
model to estimate the nonperturbative matrix elements of the 4-quark
operators \cite{du,tra} and the axial current form-factor for decuplet-octet
transition. This approach uses only one free parameter, {\em i.e.} the Skyrme
charge $e$. In order to avoid the unnecessary numerical burden, throughout
this report we use the {\it arctan Ansatz} for the Skyrme profile function $%
F(r)$ \cite{MPSU3,dia}, which allows to calculate the pertinent overlap
integrals analytically with accuracy of the order of a few \% with respect
to the exact numerical results.

It is well known that the nonleptonic weak decays of baryons can be
reasonably well described in the framework of the Standard Model \cite
{ga,do2}. The starting point in such an analysis is the effective weak
Hamiltonian in the form of the current $\otimes $ current interaction,
enhanced by Quantum Chromodynamics (QCD), {\em i.e.} obtained by
integrating out the heavy-quark and the $W$-boson fields,
\begin{equation}
H_{w}^{eff}(\Delta S=1)=\sqrt{2}G_{F}V_{ud}^{\ast
}V_{us}\sum_{i=1}^{4}c_{i}O_{i},  \label{Heff}
\end{equation}
where $G_{F}$ is the Fermi constant and $V_{ud}^{\ast }V_{us}$ are the
Cabbibo-Kobayashi-Maskawa matrix elements. The coefficients $c_{i}$ are the
well known Wilson coefficients \cite{do2}, most recently evaluated in Ref.
\cite{bur} and the $O_{i}$ are the familiar 4-quark operators
\cite{do2,bur}. For the purpose of this work we neglect the so called Penguin
operators since their contributions are proven to be small \cite{do2,bur}.
We are using the Wilson coefficients from Ref.\cite{do2}: $%
c_{1}=-1.90-0.61\zeta \;\;,c_{2}=0.14+0.020\zeta
\;\;,c_{3}=c_{4}/5\;\;,c_{4}=0.49+0.005\zeta $, with $\zeta =V_{td}^{\ast
}V_{ts}/V_{ud}^{\ast }V_{us}$. Without QCD-short distance corrections, the
Wilson coefficients would have the following values: $c_{1}=-1,\;c_{2}=1/5,%
\;c_{3}=2/15,\;c_{4}=2/3\,$. In this paper we simply consider both
possibilities and compare the resulting amplitudes.

The techniques used to describe nonleptonic $\Omega ^{-}$ decays ($%
3/2^{+}\rightarrow 1/2^{+}+0^{-}$ reactions involve only p- and d-waves) are
known as a modified current-algebra (CA) approach. The general form
of the decay amplitude reads:
\begin{eqnarray}
<\!\pi (q)B^{\prime }(p^{\prime })|H_{w}^{eff}|B(p)\!>\! &=&{\cal {\overline U}}
(p^{\prime })[{\cal B}+\gamma _{5}{\cal C}]q^{\mu }{\cal W}_{\mu }(p) .
\label{P+S}
\end{eqnarray}
Here ${\cal U}(p^{\prime })$ denotes regular spinor while and ${\cal W}_{\mu
}(p)$ is the Rarita-Schwinger spinor. The parity-conserving amplitudes $%
{\cal B}$ correspond to the p-wave and parity-violating amplitudes ${\cal C}$
correspond to the d-wave $\Omega ^{-}$ decays, respectively.

The decay probability $\Gamma (3/2^{+}\rightarrow 1/2^{+}+0^{-})$ reads:
\begin{eqnarray}
\Gamma &=&\frac{|{\bf p}^{\prime }|^{3}}{12\pi m_{\Omega }}\left[ (E^{\prime
}+m_{f})|{\cal B}|^{2}+(E^{\prime }-m_{f})|{\cal C}|^{2}\right] ,  \nonumber
\\
|{\bf p}^{\prime }|^{2} &=&\left[ (m_{\Omega }^{2}-m_{f}^{2}+m_{\phi
}^{2})/2m_{\Omega }\right] ^{2}-m_{\phi }^{2},  \nonumber \\
E^{\prime } &=&(m_{\Omega }^{2}+m_{f}^{2}-m_{\phi }^{2})/2m_{\Omega }.
\label{Gamma}
\end{eqnarray}
Here $m_{f}$ denotes the final baryon mass and $m_{\phi }$ is the mass of
the emitted meson. Obviously the parity-violating amplitude ${\cal C}$
(d-wave) is multiplied in $\Gamma $ by an unfavorable factor $(E^{\prime
}-m_{f})$. In our calculation framework, the amplitude ${\cal C}$ can
receive contributions only from the pole diagrams. As these pole diagrams
contain negative-parity $1/2^{-}$ and $3/2^{-}$ baryon resonances, the sum
of baryon masses instead of the difference appears in denominator.
Parity-violating amplitudes are thus suppressed by a large factor. If the
vertices in both ${\cal B}$ and ${\cal C}$ pole contributions are of the
same order of magnitude, then the decay of $\Omega ^{-}$ is almost
parity-conserving. This conclusion is the same as the one drawn in
Ref.\cite{fin}.

We calculate the parity-conserving p-wave $\Omega ^{-}$ decay amplitudes $%
{\cal B}$ using the so-called tree-diagram approximation at the particle
level. This means that in this work we take into account all possible tree
diagrams involving baryons and mesons, {\em i.e.} factorizable and pole
diagrams.

All ${\cal B}$ amplitudes receive contributions from the pole diagrams, as
shown in Fig.1, and they are as follows:
\begin{eqnarray}
B_{{\cal P}}(\Omega _{K}) &=&\frac{g_{\Omega ^{-}K^{-}\Xi ^{0}}a_{\Lambda
\Xi _{0}}}{m_{\Xi ^{0}}-m_{\Lambda }}-\frac{g_{\Xi ^{\ast -}K^{-}\Lambda
}a_{\Xi ^{\ast -}\Omega ^{-}}}{m_{\Omega ^{-}}-m_{\Xi ^{\ast -}}},  \nonumber
\\
B_{{\cal P}}(\Omega _{-}^{-}) &=&-\frac{g_{\Xi ^{0}\Xi ^{\ast -}\pi
^{-}}a_{\Xi ^{\ast -}\Omega ^{-}}}{m_{\Omega ^{-}}-m_{\Xi ^{\ast -}}},
\nonumber \\
B_{{\cal P}}(\Omega _{0}^{-}) & = & -\frac{g_{\Xi ^{-}\Xi ^{\ast -}\pi
^{0}}a_{\Xi ^{\ast -}\Omega ^{-}}}{m_{\Omega ^{-}}-m_{\Xi ^{\ast -}}}.
\label{poles}
\end{eqnarray}
\begin{figure}[bth]
\centerline{\epsfig{file=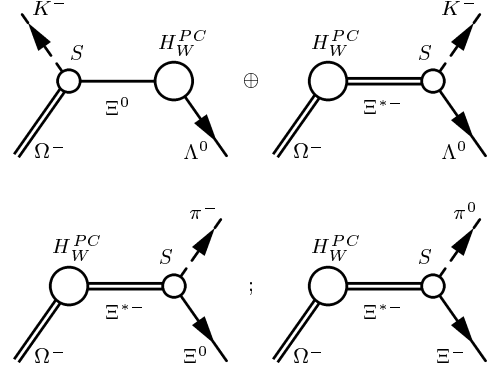,silent=}}
\caption{Pole diagrams. The double lines are $3/2^+$ resonances. The full
lines are hyperons and the dashed lines are mesons. Weak and strong vertices
are indicated.}
\end{figure}

The decuplet-octet-meson strong coupling constants determined from the
experimental value of the decay rate $\Delta ^{++}\rightarrow p\pi ^{+}$ and
by using SU(3) relations are as follows:
\begin{eqnarray}
g_{\Delta ^{++}\pi ^{+}p} &=&g_{\Xi ^{0}\Omega ^{-}K^{-}}=-g_{\Omega ^{-}\Xi
^{-}\pi ^{0}}=\sqrt{2}g_{\Lambda ^{0}\Xi ^{\ast -}K^{-}}  \nonumber \\
=\sqrt{3}g_{\Xi ^{0}\Xi ^{\ast -}\pi ^{-}} &=&\sqrt{6}g_{\Xi ^{-}\Xi ^{\ast
-}\pi ^{0}}=15.75\text{GeV}^{-1}.  \label{mesg}
\end{eqnarray}
Note that the strong couplings $g_{\Delta N\pi }$ and $g_{NN\pi }$ are related
by $g_{\Delta N\pi }=\frac{3}{2}g_{NN\pi }$, which follows from the $1/N_{c}$
expansion without other assumptions\cite{adk}. This relation is in excellent
agreement with experiment and with the evaluation of $g_{NN\pi }$ and
$g_{\Delta N\pi }$ in the Skyrme model\cite{adk}. Note also that $g_{NN\pi }$ and
$g_A^{NN}$ are related by the Goldberger-Treiman relation,
which is also true in the Skyrme model.
So altogether we can reproduce the experimental value of $g_{\Delta N\pi }$ if we use the
Skyrme model parameters which reproduce $g_A^{NN}$.
However, consistently following the approach of Ref. \cite{du}, we are
using the Skyrme model to estimate only the unknown matrix elements.
Therefore, it is natural to use the experimental value
$g_{\Delta ^{++}\pi ^{+}p}=15.75$ GeV$^{-1}$
in the calculations of baryon poles (\ref{poles}).

The baryon-pole $B_{{\cal P}}$ amplitudes contain weak matrix elements
defined as $a_{BB^{\prime}} = {\cal G} <B^{\prime}|%
\sum_{i=1}^{4}c_{i}O_{i}^{PC}|B>$, were the constant ${\cal G} = \sqrt{2}G_F
V^*_{ud}V_{us}$. The important part of $a_{BB^{\prime}}$ are 4-quark
operator matrix elements, which are nonperturbative quantities. This is the
first point of this work at which the Skyrme model is used.

The factorizable contributions to p-waves (only for ($\Omega^-_{-,0}$) amplitudes), are calculated by
inserting the vacuum states; it is therefore a factorized product of two
current matrix elements, where the octet-decuplet matrix element of the
axial-vector current reads:
\[
<\Xi(p^{\prime })|A^{\mu }|\Omega ^{-}(p)>=g_{A}^{\Xi \Omega }(q^{2})
{\cal {\overline U}}_{\Xi }(p^{\prime }){\cal W}_{\Omega }^{\mu }(p).
\]
Summing over all factorizable contributions gives the following expressions
for the amplitudes:
\begin{eqnarray}
B_{{\cal S}}(\Omega _{-}^{-})&=&\frac{1}{3\sqrt 2}{\cal G}f_{\pi }g_{A}^{\Xi
^{0}\Omega ^{-}}\left[ c_{1}-2(c_{2}+c_{3}+c_{4})\right], \nonumber \\
B_{{\cal S}}(\Omega _{0}^{-})&=&\frac{1}{6}{\cal G}f_{\pi }g_{A}^{\Xi
^{-}\Omega ^{-}}\left[ c_{1}-2(c_{2}+c_{3}-2c_{4})\right] . \label{sed}
\end{eqnarray}
The $g_{A}^{\Xi \Omega }$ represents the form-factor of the spatial
component of the axial-vector current between $\Xi ^{{0 \choose -}}$
and $\Omega ^{-}$ states. This is the second point of this paper where the
Skyrme model enters the calculation. Note that the 
$|B_{{\cal S}}(\Omega _{0}^{-})| \ll |B_{{\cal S}}(\Omega _{-}^{-})|$ due to the helicity suppression
enhanced by the QCD corrections.

Total theoretical amplitudes are:
\begin{equation}
{\cal B}_{\rm theory}(q^2 )=B_{{\cal P}}(q^2 )+B_{{\cal S}}(q^2 ),
\label{Bth}
\end{equation}
were the relative signs between pole and factorizable contributions
are determined via SU(3) and the generalized Goldberger-Treiman relations.

In order to estimate the matrix elements entering Eqs.(\ref{poles})
 and (\ref{sed}), we take
the SU(3) extended Skyrme Lagrangian \cite{wei,tra,sco}:
\begin{equation}
{\cal L}={\cal L}_{\sigma }+{\cal L}_{Sk}+{\cal L}_{SB}+{\cal L}_{WZ},
\label{Lagr}
\end{equation}
where ${\cal L}_{\sigma }$, ${\cal L}_{Sk}$, ${\cal L}_{SB}$, and ${\cal L}%
_{WZ}$ denote the $\sigma $-model, Skyrme, symmetry breaking (SB), and
Wess-Zumino (WZ) terms, respectively. Their explicit form can be found in
Refs.\cite{wei,sco}. In this work we will use the SU(3) extended Lagrangian
(\ref{Lagr}) in the limit of $f_{K}=f_{\pi }$. Then our new set of parameters ${%
\hat{x}}$, $\beta ^{\prime }$, and $\delta ^{\prime }$, determined from the
masses and decay constants of the pseudoscalar mesons\cite{wei}, is: ${\hat{x%
}}=2m_{K}^{2}/m_{\pi }^{2}-1,\;\;{\beta }^{\prime }=0,\;\;{\delta }^{\prime
}=m_{\pi }^{2}f_{\pi }^{2}/4$. Owing to the presence of the $\delta ^{\prime
}$ term in classical mass $E_{cl}$, the dimensionless size of soliton $%
x_{0}^{\prime }$ becomes a function of $e$, $f_{\pi }$, and $\delta ^{\prime
}$:
\begin{equation}
x^{\prime }{}_{0}^{2}=\frac{15}{8}\left[ 1+\sqrt{1+\frac{30\delta ^{\prime }%
}{e^{2}f_{\pi }^{4}}}\;\right] ^{-1}.
\end{equation}

Note that $\hat{x}$ is responsible for the baryon mass
splittings and the admixture of higher representations in the baryon wave
functions. However, as explained above, in this work we use the SU(3)
symmetric baryon wave functions in the spirit of the perturbative approach
to SB. Indeed, we have shown in Ref.\cite{du}, that the SB effects through
the weak operators in the decay amplitudes are very small.

Our fitting procedure is as follows. Since the coupling $f_{\pi }$ is equal
to its experimental value, the only remaining free parameter is the Skyrme
charge $e$. The value $e \approx 4$ was successfully adjusted to the mass
difference of the low-lying $1/2^{+}$ and $3/2^{+}$ baryons (Table 2.1 of
\cite{wei}). This value of $e$ was next employed to evaluate the static
properties of baryons.

For the evaluation of the nonleptonic $\Omega ^{-}$ decays, the most
important baryon static properties is the octet-decuplet mass splitting $%
\Delta $. Other important quantity is the axial coupling constant
$g_{A}^{\rm pn}$. We compute these quantities by using the {\it
arctan Ansatz} in the SU(3)
extension of the Skyrme Lagrangian (\ref{Lagr}). By fixing $\Delta $ and $%
g_{A}^{\rm pn}$ to their experimental values, we obtain $e=4.21$ and $e=3.41$,
respectively. In further calculations of the $\Omega ^{-}$ decay amplitudes,
we use the mean value $e=3.81$, like in the case of the nonleptonic hyperon
decays\cite{du}. As in the latter case this introduces approximately
15 \% uncertainty of the decay amplitudes which are dominated by
$\Phi^{SK}$ (see (\ref{oct}) below) which scales like  $1/e$.

We proceed with the computation of the axial-current form-factor
$g_{A}^{\Xi \Omega }$ in the Skyrme model using the {\it arctan
Ansatz}:
\begin{equation}
g_{A}^{\Xi ^{-}\Omega ^{-}}=g_{A}^{\Xi ^{0}\Omega ^{-}}\equiv g_{A}^{\Xi
\Omega }(x_{0}^{\prime })=\frac{2}{7}\sqrt{15}\;
g_{A}^{\rm pn}(x_{0}^{\prime }),
\label{gA}
\end{equation}
were $g_{A}^{\rm pn}(x_{0}^{\prime })$ is given in eq.(10) of Ref.\cite{du}.

Let us recapitulate results of Ref.\cite{du} where we have calculated the
matrix element of the product of two $(V-A)$ currents between the octet
states using the Clebsch-Gordon decomposition \cite{tra,sco}:
\begin{equation}
<B_{2}|{\hat{O}}^{(SK)}|B_{1}>=\Phi ^{SK}\times \sum_{R}C_{R}
\end{equation}
where $\Phi ^{SK}$ is a dynamical constant and $C_{R}$ denote the pertinent
sum of the SU(3) Clebsch-Gordon coefficients in the intermediate
representation $R$. The same holds for the WZ current. The total matrix
element is then simply a sum $<{\hat{O}}_{i}^{(SK)}+{\hat{O}}_{i}^{(WZ)}>$,
with $i=1,\ldots ,4$. The quantities $\Phi $ are given by the overlap
integrals of the profile function in Ref.\cite{du}.

For the ${\hat{O}}_{1}$ operator, $R=8_{a,s}$ or $27$. The matrix elements
read:

\begin{eqnarray}
&<&\Lambda _{1/2}^{0}|{\hat{O}}_{1}|\Xi _{1/2}^{0}>=(18.14\Phi
^{SK}-20.41\Phi ^{WZ})10^{-3}  \nonumber \\
&=&(4.86|_{SK}-0.09|_{WZ})10^{-3}\text{GeV}^{3},  \label{oct} \\
&<&\Xi _{3/2}^{\ast -}|{\hat{O}}_{1}|\Omega _{3/2}^{-}>=(-10.60\Phi
^{SK}+12.89\Phi ^{WZ})10^{-3}  \nonumber \\
&=&(-2.84|_{SK}+0.06|_{WZ})10^{-3}\text{GeV}^{3}.  \label{dec}
\end{eqnarray}
Note that we have omitted the static kaon fluctuations in all of our
computations.

Our main goal was to learn how the approach of Refs.\cite{du,ga}
in which the Skyrme model is used to estimate the unknown nonperturbative
matrix elements applies to the $\Omega ^{-}$ nonleptonic decays.
We have systematically presented all possible tree diagrams, namely
factorizable contributions, octet diagram (only for ($\Omega _{K}$) amplitude) 
and decuplet pole diagrams. Numerical results
presented in Table \ref{t:tab1} are
very encouraging. They are in satisfactory quantitative agreement with
experimental data. In Table \ref{t:tab1} both factorizable ($B_{{\cal S}%
}(q^{2})$) and pole-amplitudes ($B_{{\cal P}%
}(q^{2})$) are displayed. Comparison
of the total amplitudes (\ref{Bth}) ${\cal B}_{\rm theory}(q^{2})$
with experiment shows the following:

(a) Dynamics based on the pole-diagrams supported by the Skyrme model leads to very
good results for relative importance of various decay modes. However, all
amplitudes are too large by about a factor of two.

(b) The decay amplitude 
$\Omega _{K}$, which does not contain
factorizable contributions, has the largest pole contribution, since both
pole diagrams are of approximately the same size and contribute constructively.

(c) Short-distance corrections to the effective weak Hamiltonian are proved
to be important\cite{do2,bur}.

(d) In fact, too large values of the QCD-enhanced $B_{{\cal P}}(q^{2})$
amplitudes are actually a welcome feature. Namely, if one takes into account
the higher $\Xi (3/2^{+})$ resonances then, the poles (\ref{poles}) would flip the
signs, since their masses are larger than the $\Omega ^{-}$ mass.
The assumption that the vertices would be more or less the same like for
the ground-state poles leads to internal cancellations between
pole-diagram contributions \cite{foot,mil}. Taking into account (a) -- (c)
and Ref.\cite{mil} we expect that this
dynamical scheme would produce better agreement with experiment.

(e) We have found the octet dominance {\em i.e.} the 27-contaminations
are small.

(f) We also find the dominance of the Skyrme Lagrangian currents over the WZ
current.
For $e\approx 4$, the WZ contribution to (\ref{oct},\ref{dec}%
) is below $3\%$.

(g) Like in the nonleptonic hyperon decays, the factorizable contributions,
in the Skyrme model approach, turn out to be not important for the $\Omega
^{-}$ nonleptonic decays.

(h) Finally, we have found that the pole terms and the factorizable
contributions have opposite signs.

We conclude that in general our approach provides
the good description of the $\Omega ^{-}$ decays. Obviously, not
all details are under full control ({\em e.g.} $m_{s}$ corrections are
neglected); nevertheless, it seems the QCD-corrected weak Hamiltonian $%
H_{w}^{eff}$, together with the inclusion of other possible types of
contribution to the total amplitudes ($K $-poles and/or
factorization, higher baryon-poles, etc.) supplemented by the
Skyrme model, lead to a correct answer.

We hope that the present calculation,
taken together with the analogous calculation of the nonleptonic
hyperon decay amplitudes\cite{du,ga} will contribute to the understanding of
the nonleptonic hyperon interactions. It is also a test of the application of
the Skyrme model to the evaluation of the nonperturbative quantities like axial
coupling constants and the dimension six operator matrix elements between
the different baryon states.

This work was supported by the Ministry of Science and Technology of the
Republic of Croatia under the contract No. 00980102. M.P. was supported by
the Polish KBN Grant PB~2~P03B~{\-}019~17.

\renewcommand{\arraystretch}{1.5}
\begin{table}[tbp]
\caption{The p-wave (${\cal B}$) nonleptonic $\Omega^-$ decay amplitudes.
Choices (off, on) correspond to the amplitudes without and with inclusion of
short-distance corrections, respectively. For $\Omega_K(\Omega^-_{0,-})$
amplitude $q^2=m^2_K(m^2_{\protect\pi})$.}
\label{t:tab1}
\begin{tabular}{c|ccccccccc}
Amplitude $(10^{-6}{\rm GeV}^{-1}) $ & & & &$(\Omega_K) $ &  & $(\Omega^-_-) $
&  & $(\Omega^-_0) $ &  \\ \hline\hline
$B^{(1/2^+)}_{{\cal P}}(q^2)$ &  & off  &  & $2.67$ &  & $0$ &  & $0$ &
\\
&  & on  &  & $5.51$ &  & $0$ &  & $0$ &  \\ \hline
$B^{(3/2^+)}_{{\cal P}}(q^2)$ &  & off  &  & $1.57$ &  & $1.28$ &  & $0.91$
&  \\
&  & on  &  & $3.14$ &  & $2.56$ &  & $1.81$ &  \\ \hline
$B_{{\cal S}}(q^2) $ &  & off  &  & $0 $ &  & $-0.27$ &  & $-0.06$ &  \\
&  & on  &  & $0 $ &  & $-0.30$ &  & $0.03$ &  \\ \hline
${\cal B}_{\rm theory}(q^2) $ &  & off &  & $4.24$ &  & $1.01$ &  & $0.85$ &
\\
&  & on  &  & $8.65$ &  & $2.26$ &  & $1.84$ &  \\ \hline
${\cal B}_{\rm exp}(10^{-6}{\rm GeV}^{-1})$\cite{rpp} &  &  &  & $4.02 $ &  & $1.35 $
&  & $0.80 $ &
\end{tabular}
\end{table}


\begin{references}
\vspace{-1cm}
\bibitem{du}  G. Duplan\v{c}i{\'{c}}, H. Pa\v{s}agi{\'{c}}, M. Prasza{\l }%
owicz and J. Trampeti\'{c}, Phys. Rev. {\bf D64}, 097502 (2001); hep-ph/0104281.

\bibitem{rpp}  Rev. of Part. Prop., Eur. Phys. J. {\bf C15}, (2000).

\bibitem{sky}  T.H.R. Skyrme, Proc. Roy. Soc. {\bf A260}, 127 (1961).

\bibitem{adk}  G. Adkins, C.R. Nappi and E. Witten, Nucl. Phys. {\bf B228},
552 (1983).

\bibitem{gua}  E. Guadagnini, Nucl. Phys. {\bf B236}, 35 (1984).


\bibitem{NMPJW}  P.O. Mazur, M.A. Nowak and M. Prasza{\l }owicz, Phys. Lett.
{\bf 147B}, 137 (1984); S. Jain and S. Wadia, Nucl. Phys. {\bf B258}, 713
(1985).

\bibitem{MPSU3}  M. Prasza{\l }owicz, Phys. Lett. {\bf 158B}, 264 (1985);
Acta. Phys. Pol. {\bf B22}, 525 (1991).

\bibitem{wei}  H. Weigel, Int. J. Mod. Phys. {\bf A11}, 2419 (1996); J.
Schechter and H. Weigel, hep-ph/9907554 (1999).


\bibitem{ga}  H. Gali{\'{c}}, D. Tadi{\'{c}} and J. Trampeti{\'{c}}, Phys.
Lett. {\bf 89B}, 249 (1980); D. Tadi{\'{c}} and J. Trampeti{\'{c}}, Phys.
Rev. {\bf D23}, 144 (1981).

\bibitem{tra}  J. Trampeti\'{c}, Phys. Lett. {\bf 144B}, 250 (1984); M.
Prasza{\l }owicz and J. Trampeti\'{c}, Phys. Lett. {\bf 161B}, 169 (1985);
Fizika {\bf 18}, 391 (1986).

\bibitem{dia}  D.I. Diakonov, V.Yu. Petrov and M. Prasza{\l }owicz, Nucl.
Phys. {\bf B323}, 53 (1989).

\bibitem{do2}  J.F. Donoghue, H. Golowich and B.R. Holstein, {\it Dynamics
of the Standard Model} (Cambridge Univ. Press, 1994).

\bibitem{bur}  A.J. Buras, Les Houches Lectures; hep-ph/9806471.

\bibitem{fin}  J. Finjord, Phys. Lett. {\bf 76B}, 116 (1978).

\bibitem{sco}  N.N. Scoccola, Phys. Lett. {\bf B428}, 8 (1998); D. Dumm, A.
Garcia and N. Scoccola, Phys. Rev. {\bf D62}, 014001 (2000).

\bibitem{foot} The paragraph, 2.3. of Ref.\cite{mil} explicitely shows such internal
cancellations between pole-diagrams on the example of the p-wave
$\Sigma^+ \rightarrow n \pi^+$ decay.

\bibitem{mil} M. Milo{\v{s}}evi{\'{c}},  D. Tadi{\'{c}} and J. Trampeti{\'{c}},
Nucl. Phys. {\bf B207}, 461 (1982).
\end{references}
\end{document}